%% file: main_JINST.tex
\documentclass[a4paper,11pt]{article}
\usepackage{jinstpub} % for details on the use of the package, please see the JINST-author-manual
\usepackage{lineno}

\usepackage{graphicx}
\usepackage{dcolumn}
\usepackage{bm}
\usepackage{siunitx}
\usepackage{hyperref}
\usepackage{subcaption}
\usepackage{orcidlink}
\usepackage{booktabs}
\usepackage{csquotes}
\usepackage{textcase}

\graphicspath{{Figs/}}
\title{Performance of FBK VUV-HD3 and HPK VUV4 SiPMs in the Light-only Liquid Xenon (LoLX) Detector}

\newcommand{\afSFU}{Department of Physics, Simon Fraser University, 8888 University Drive, Burnaby, BC, V5A 1S6, Canada}
\newcommand{\afQueens}{Department of Physics, Engineering Physics and Astronomy, Queen's University, 64 Bader Lane, Kingston, ON, K7L 3N6, Canada}
\newcommand{\afTRIUMF}{TRIUMF, 4004 Wesbrook Mall, Vancouver, BC, V6T 2A3, Canada}
\newcommand{\affilUBC}{Department of Physics and Astronomy, The University of British Columbia,
325 - 6224 Agricultural Road,
Vancouver, BC V6T 1Z1
}
\newcommand{\affilUBCforestry}{Faculty of Forestry, The University of British Columbia,
2424 Main Mall,
Vancouver, BC V6T 1Z4
}
\newcommand{\affilMcGill}{Department of Physics, McGill University, 3600 University Street, Montreal, QC, H3A 2T8, Canada}

\newcommand{\affilsherbrook}{Faculté des Sciences, 2500 Bd de l'Université, Sherbrooke, Quebec J1K 2R1}

\newcommand{\afDrexel}{Department of Physics, Drexel University, Philadelphia, PA 19104, USA}
\newcommand{\afOAU}{Division of Applied Nuclear Science and Technology, Centre for Energy Research and Development, Obafemi Awolowo University, Ile-Ife 220005, Nigeria}

\affiliation[a]{\afSFU}
\affiliation[b]{\afTRIUMF}
\affiliation[c]{\affilMcGill}
\affiliation[d]{\afQueens}
\affiliation[e]{\affilUBC}
\affiliation[f]{\affilUBCforestry}
\affiliation[g]{\affilsherbrook}
\affiliation[h]{\afDrexel}
\affiliation[i]{\afOAU}

\author[a,b,1]{Xiang Li~\orcidlink{0009-0006-0322-3017}}
\note{Corresponding author.}

\author[b, c]{David Gallacher~\orcidlink{0000-0002-9395-0560}} 

\author[b,f]{Stephanie Bron}

\author[c]{Thomas Brunner~\orcidlink{0000-0002-3131-8148}}

\author[b,d]{Austin de St Croix~\orcidlink{0000-0002-6517-420X}}

\author[c]{Frédéric Girard~\orcidlink{0000-0003-0537-6296}}

\author[b,e]{Colin Hempel}

\author[h, i]{Mouftahou Bakary Latif~\orcidlink{0000-0002-9326-4456}}

\author[c]{Simon Lavoie~\orcidlink{0009-0000-6336-511X}}

\author[b,c,e]{Chloé Malbrunot~\orcidlink{0000-0001-6193-6601}}

\author[b]{Fabrice Retière}

\author[g]{Marc-André Tétrault ~\orcidlink{0000-0003-2887-4335}}

\author[b]{Lei Wang}

\emailAdd{xla405@sfu.ca}

\input{section-abstract}


\begin{document}
\maketitle
\flushbottom

%--------------------------------------------------
\input{section-Intro}

\input{section-detector}

\input{section-MeasurementProcedure}
\input{section-sim_fram}
\section{Results and Discussion}\label{sec:results}

    \input{section_SiPMCharge}  %% subsection 1
    \input{section-sim_results}  %% subsection 2
\input{section-Discussion}
\input{section-conclusion}

%--------------------------------------------------
\acknowledgments
This research was undertaken thanks in part to funding from the Canada First Research Excellence Fund through the Arthur B. McDonald Astroparticle Physics Research Institute, with support from the Natural Sciences and Engineering Research Council of Canada (NSERC), the Fonds de Recherche du Québec (FRQ), and the Canada Foundation for Innovation (CFI). We wish to thank the TRIUMF Science and Technology Group for their invaluable support in the construction of the detector. We also thank the Digital Research Alliance of Canada and Calcul Québec for providing the computing resources required to undertake this work. 
% We gratefully acknowledge the support of Canadian taxpayers

% Bibliography

%% [A] Recommended: using JHEP.bst file
\bibliographystyle{JHEP}
\bibliography{
reference
}
%% or
%% [B] Manual formatting (see below)
%% (i) We suggest to always provide author, title and journal data or doi:
%% in short all the informations that clearly identify a document.
%% (ii) please avoid comments such as "For a review'', "For some examples",
%% "and references therein" or move them in the text. In general, please leave only references in the bibliography and move all
%% accessory text in footnotes.
%% (iii) Also, please have only one work for each \bibitem.

% \begin{thebibliography}{99}

% \end{thebibliography}
\end{document}

%% file: section-abstract.tex
\abstract{
Silicon Photomultipliers (SiPMs) are a critical technology for the next
generation of rare-event search experiments using liquid xenon (LXe). LXe scintillation is emitted primarily in the vacuum-ultraviolet (VUV), requiring photosensors with high VUV detection efficiency under cryogenic operating conditions. Direct characterization of VUV-sensitive SiPMs in LXe is therefore important for informing detector design and comparing device response. This work presents a direct comparison of Fondazione Bruno Kessler (FBK) VUV-HD3 and Hamamatsu (HPK) VUV4
SiPMs operated simultaneously within the Light-only Liquid Xenon (LoLX)
detector. Using data collected with gamma sources placed outside the detector, we characterized the relative
performance of these photosensors. Our analysis reveals that under these operating conditions, the HPK SiPMs observe 33–38\% less light than the FBK devices, a larger difference than predicted by standard PDE models in vacuum measurements. We show that this discrepancy is resolved by an angular and wavelength dependent PDE model incorporating surface shadowing effects into our optical simulation, which then accurately reproduces the experimental data. This finding has significant implications for the selection and implementation of photosensors in future large-scale LXe detectors.
}

%% file: section-Intro.tex
\section{Introduction}\label{sec:intro}

Liquid xenon (LXe) detectors have emerged as a leading technology for rare-event searches in fundamental physics, with applications ranging from dark matter detection experiments \cite{LZ:2019sgr, PandaX:2024syk, XENON:2024wpa} and neutrinoless double beta decay searches \cite{EXO-200:2019rkq}, to tests of lepton flavor universality in the PIONEER experiment \cite{PIONEER:2025}. LXe also has applications in medical technology, such as PET imaging \cite{LoLX:2023vyw, Romo-Luque:2019ohl}.
These detectors exploit xenon's excellent scintillation properties, producing vacuum ultraviolet (VUV) photons centred at approximately \SI{175}{nm}, with a high scintillation light yield, proportional to the energy of incident radiation \cite{Fujii:2015qqq}. 

The performance of LXe detectors critically depends on the photon detection efficiency (PDE) and characteristics of the photosensors used to collect the VUV scintillation light. Traditionally, photomultiplier tubes (PMTs) have been the standard choice for many xenon-based experiments \cite{LZ:2019sgr, PandaX:2024syk, XENON:2024wpa} due to their well-understood performance. However, silicon photomultipliers (SiPMs) have gained attention as promising alternatives, offering several potential advantages including compact form factors, low operating voltages, insensitivity to magnetic fields, low-background instrumentation, and excellent single-photoelectron (SPE) resolution \cite{Acerbi:2019qgp, Rogers:2024cvk}.

Different SiPM manufacturers employ various approaches to VUV sensitivity optimization, resulting in potentially significant performance differences in liquid xenon applications. Two prominent manufacturers of VUV-sensitive SiPMs are Fondazione Bruno Kessler (FBK) \cite{FBK} and Hamamatsu Photonics (HPK) \cite{HPK}. The relative performance of these two manufacturers under actual liquid xenon operating conditions has important implications for the design and optimization of future xenon detectors.

Direct comparative studies of SiPM technologies in LXe remain limited. While measurements of reflectivity for HPK and FBK devices in LXe \cite{nEXO:2021uxc} and  angular dependence of relative PDE for HPK sensors \cite{nEXO:2019jhg} exist, equivalent angular PDE data for FBK devices are currently unavailable. Furthermore, reported HPK performance metrics are inconsistent: vacuum measurements suggest a PDE of approximately 20\% \cite{nEXO:2022yey}, whereas in-situ LXe measurements by the MEG-II collaboration indicate a maximum efficiency of only 13\% prior to radiation exposure \cite{MEGII:2022klg}.

In this work, we address these disparities by presenting a detailed comparison of FBK VUV-HD3 and HPK VUV4 SiPMs operated simultaneously within the upgraded Light-only Liquid Xenon detector (LoLX 2). This system provides a controlled environment for evaluating both sensor types under identical conditions. We employ laser calibration and external gamma-ray sources to characterize their relative photon detection, comparing the results against Monte Carlo simulations. By incorporating a detailed optical model \cite{Croix:2025axo} that captures the angular dependence of photon detection efficiency, we aim to resolve the observed efficiency discrepancies and establish robust performance benchmarks for future LXe detectors.

%% file: section-detector.tex
%--------------------------------------------------
\section{Light-only-Liquid Xenon 2 experiment}\label{sec:setup}

\begin{figure}
    \centering
    \includegraphics[width=0.5\linewidth]{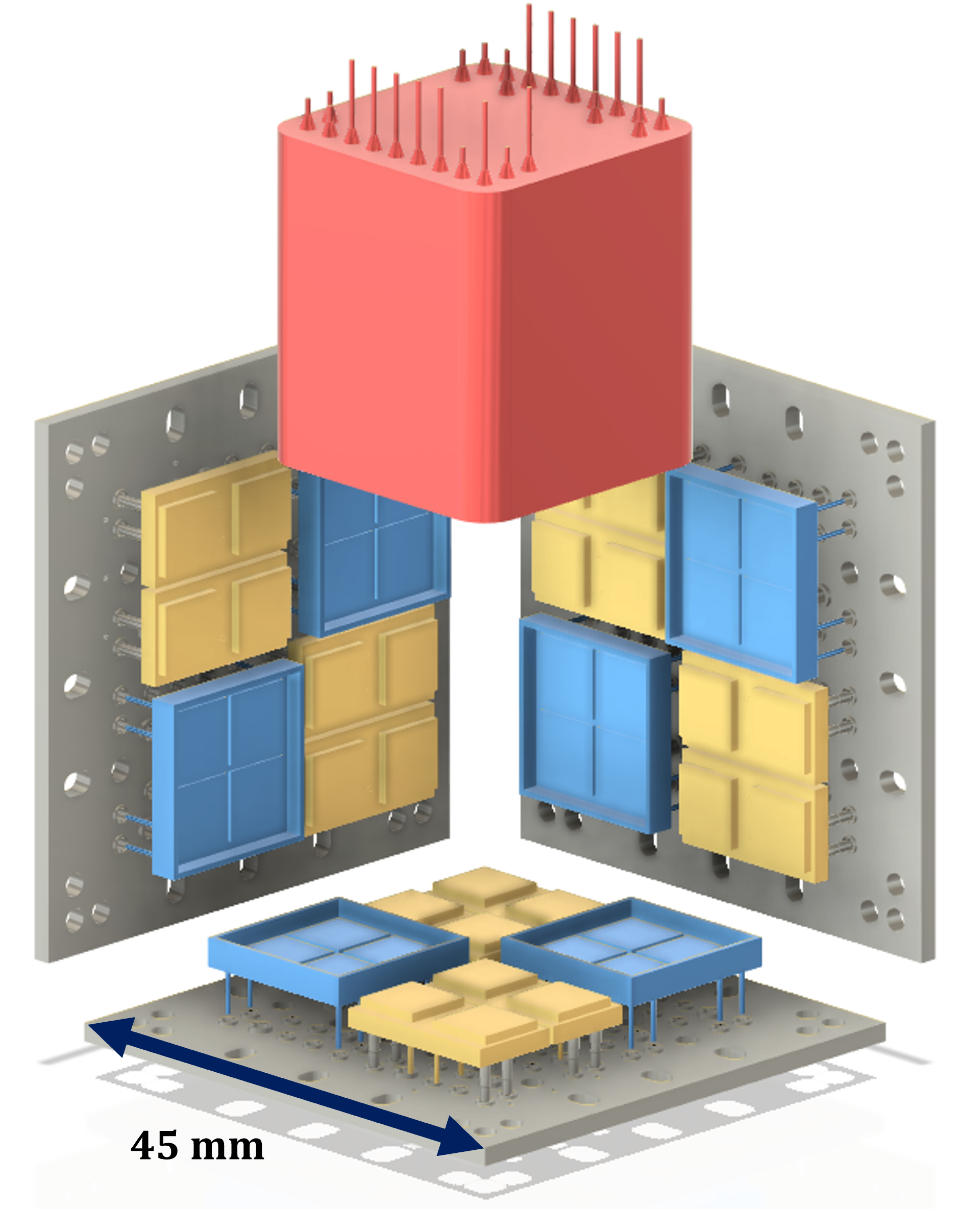}
    \caption{Three-dimensional CAD rendering of the LoLX 2 detector's photosensor configuration. The 4 cm cubic active volume is filled with liquid xenon and instrumented with three types of photosensors: Hamamatsu VUV4 SiPMs (blue), FBK VUV HD3 SiPMs (yellow), and a centrally located Hamamatsu R8520-406 SEL photomultiplier tube (PMT, red) at the top. The SiPM tiles on the two front-facing side panels have been omitted from the rendering to improve visibility of the interior photosensor layout.}
    \label{fig:geo}
\end{figure}

 LoLX 2 is a \SI{4}{\centi\meter} cubic detector immersed in approximately \SI{5}{\kilo\gram} of liquid xenon (LXe), serving as the successor to LoLX 1 \cite{lolx_ext_2025}. The new cubic design is modular, with the detector body built from PCB tiles mounted to stainless steel corner brackets. Different types of SiPMs can be mounted to each tile for comparative studies and a PMT is mounted in place of the upper tile. Optical filters can also be installed in front of each face, as is foreseen for future studies of Cherenkov light \cite{LoLX:2023vyw}. Each tile holds 16~SiPMs in a $4\times4$ configuration, giving a total of 80~channels distributed across the five faces: 40 FBK VUV HD3 and 40 HPK VUV4 units, with a VUV-sensitive Hamamatsu PMT (R8520-406 SEL PMT) positioned on the sixth face located on the top of the cube. Each face of the cube is populated with both HPK and FBK devices. Figure~\ref{fig:geo} illustrates the three-dimensional arrangement of these photosensors. Crucially, the HPK VUV4 units feature a recessed window within a ceramic package, while the FBK VUV-HD3 units have a flat topology; details of both geometries are visible in Figure~\ref{fig:geo}. 

During LXe filling, high-pressure xenon gas is first routed through a pressure regulator and reduced to a pressure of $\sim$\SI{200}{kPa}. The xenon is purified before its liquefaction in the cryostat by passing it through a heated zirconium-alloy getter (MonoTorr PS3-MT3)~\cite{purgasproduct2025} and then through an ambient-temperature SAES 902 inline purifier~\cite{microtorr2025}. This purification removes impurities that can quench scintillation light production and attenuate scintillation photons during transport. During filling, the pressure in the inner vessel is kept at $\sim$\SI{100}{kPa} by automatically controlling the filling speed with an MKS GM50A-series mass flow controller~\cite{mks2025}. The cryostat is maintained at the xenon condensation point of \SI{165}{\kelvin} ($\sim$~\SI{100}{kPa}) with a SHI CH-104 cryocooler~\cite{shi2025} controlled by a Lakeshore model 350 cryogenic PID temperature controller~\cite{lakeshore2025}.
%by an open-loop, liquid-nitrogen cooling circuit with flow control.  

 The experiment utilizes the WaveDAQ (WDAQ) digitizer system from the MEG-II experiment \cite{MEGII:2019nmv} as part of the LoLX 2 upgrade. With high sampling rates of 1-5 GS/s, this system provides improved timing resolution for temporal studies of light emission. The WDAQ maintains a \SI{2}{\volt} dynamic range with 12-bit resolution suitable for single-photon detection, but operates with inherent trade-offs, including a constrained 1024-entry waveform window and higher baseline noise due to increased input bandwidth. The system is integrated within the MIDAS \cite{MIDAS_TRIUMF} framework and includes specialized timing and voltage calibration, and board synchronization procedures that are performed before each data collection period.  

\begin{figure}
    \centering
    \includegraphics[width=\linewidth]{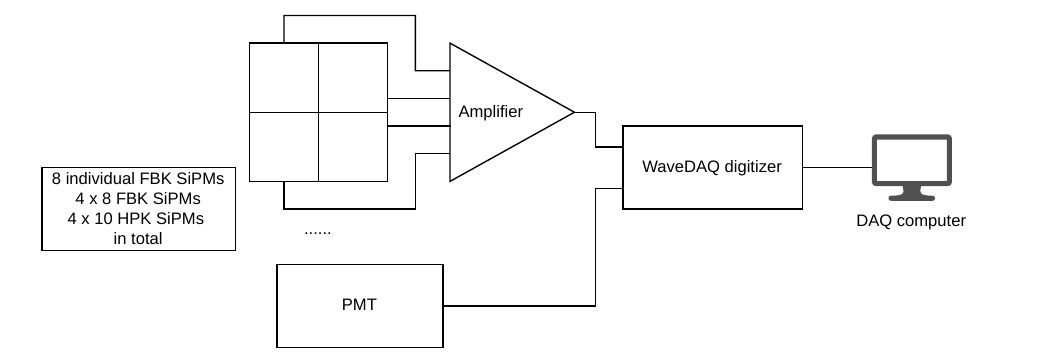}
    \caption{Simplified LoLX 2 readout and control diagram. SiPM signals were routed through amplifier boards before being digitized by the WaveDAQ system and recorded by the DAQ computer.}
    \label{fig:diagram}
\end{figure}

The SiPM signals are carried over coaxial cables and a vacuum coaxial feedthrough to custom amplifier boards. The amplified signals are subsequently carried to the WDAQ for digitization before being transmitted to DAQ PC for event recording through custom MIDAS frontend software. Eight FBK SiPMs on the bottom tile were read out individually, while the remaining FBK and HPK SiPMs were summed in groups of four at the amplifier stage. The DAQ computer employed the MIDAS framework, including custom software to interface with the amplifier boards which control the SiPM bias voltages. Additional utilities include: controlling PMT high-voltage, slow control for pressure and temperature monitoring, and data taking run control. A simplified connection diagram is shown in Figure~\ref{fig:diagram}.

%% file: section-MeasurementProcedure.tex
% \subsection{Measurement procedure}\label{sec:calib}
\subsection{Detector Commissioning and Data Acquisition}\label{sec:calib}

This section presents measurements obtained during the LoLX 2 commissioning run conducted in August 2023. Data were collected over a 3-day period with the detector filled with liquid xenon. During this campaign, background and photosensor calibration measurements were performed using an externally driven visible laser, together
with external gamma-source runs for detector calibration and characterization. The SiPMs were operated at a nominal overvoltage of \(V_{\mathrm{OV}}=3~\mathrm{V}\) in this work. This operating point was chosen as a compromise between photon-detection performance and correlated avalanche noise. At lower overvoltage, the reduced gain and photon detection efficiency limit the detected scintillation signal, whereas operation at higher overvoltage increases correlated avalanches, which can degrade the energy resolution. An overvoltage near \(3~\mathrm{V}\) has been shown to provide suitable performance for the nEXO energy-resolution requirements \cite{nEXO:2022yey}. Due to high noise interference on a specific readout channel, a subset of channels was disabled for the data taking, leaving 34 FBK and 35 HPK devices fully operational (out of 40 installed per type). The PMT was operated at a fixed gain with a \SI{750}{\volt} bias.

Each event is processed with a channel-wise pulse-finding algorithm that records every primary pulse and any resolved substructure. To compensate for minor temperature-driven gain drifts observed between runs, SPE laser calibrations are performed on a run-by-run basis. The resulting gains are used to set channel-specific analysis thresholds.

In this calibration, the number of photoelectrons (PEs) detected per channel within a fixed time window follows a Poisson distribution. Events with a waveform amplitude below the pulse-finder threshold are classified as pedestal events and assigned zero charge and pulse height. This pedestal-counting technique follows the procedure developed by the DEAP collaboration \cite{DEAP:2017fgw} and provides a simple, model-independent gain calibration for both SiPMs and PMTs. A key advantage of this method is its insensitivity to excess charge from correlated avalanches in SiPMs, as it relies on the probability of detecting zero PE rather than on the signal's magnitude. The probability of these correlated avalanches was extracted per channel during this calibration and subsequently used to correct the integrated charge of the gamma scintillation events.

% %--------------------------------------------------
\subsection{Gamma Source Measurements}
Two external gamma-ray source datasets were studied to characterize the SiPM performance: \textsuperscript{133}Ba with a primary gamma emission around \SI{356}{keV} and \textsuperscript{137}Cs emitting \SI{662}{keV} gamma rays \cite{NNDC}. The sources were positioned outside the detector, allowing gamma rays to penetrate the cryostat and deposit energy in the active LXe volume. The resulting scintillation light was measured by the SiPMs. Gamma events were triggered by the top PMT. When the PMT signal crossed the \SI{-30}{mV} trigger threshold at the WDAQ input, the WDAQ recorded the corresponding waveform window for both the PMT and SiPM channels. This trigger condition was based only on the PMT signal, ensuring that event selection was independent of the relative signal amplitudes measured by the different SiPM types.

A key challenge arises from the spatial distribution of gamma-ray energy depositions throughout the detector volume. This effect introduces strongly position-dependent variations in the detected photon count for any given SiPM, driven by solid angle coverage. These variations degrade the uniformity of the light collection, an effect which is accounted for in our comparative analysis in Section~\ref{sec:results}. For high energy events, or those close to a device, a sufficiently large number of photons is detected, saturating the digitizer. The digitizer operates over a \SI{-1}{\volt} to \SI{1}{\volt} range (a 2 V dynamic range) and features an automatic baseline at \SI{0}{\volt}. For a fraction of gamma events, the signal amplitude exceeds this range, so the affected waveform samples are clipped at the digitizer limit.
The \SI{-1}{\volt} saturation limit corresponds to approximately \SI{280}{PE} for FBK and \SI{130}{PE} for HPK devices. The different PE-equivalent saturation limits arise from the larger measured SPE waveform gain of the HPK channels. As a result, the same \(-1~\mathrm{V}\) digitizer limit is reached with fewer reconstructed PE for HPK than for FBK devices.
Approximately 30\% of events contain at least one saturated channel, though the mean number of affected channels per event remains low at ~0.4-0.5 channels. Per-channel saturation rates range from 0.1\% to 9\%, with higher-energy $^{137}$Cs events showing slightly elevated saturation compared to $^{133}$Ba.
To address this saturation for gamma calibration measurements, we employed a pulse fitting saturation correction algorithm in post-analysis. It is important to note that this saturation is an instrumental effect of the digitizer's dynamic range, as the SiPMs themselves operate well below their intrinsic saturation point. 

\begin{figure}
    \centering
    \includegraphics[width=0.95\linewidth]{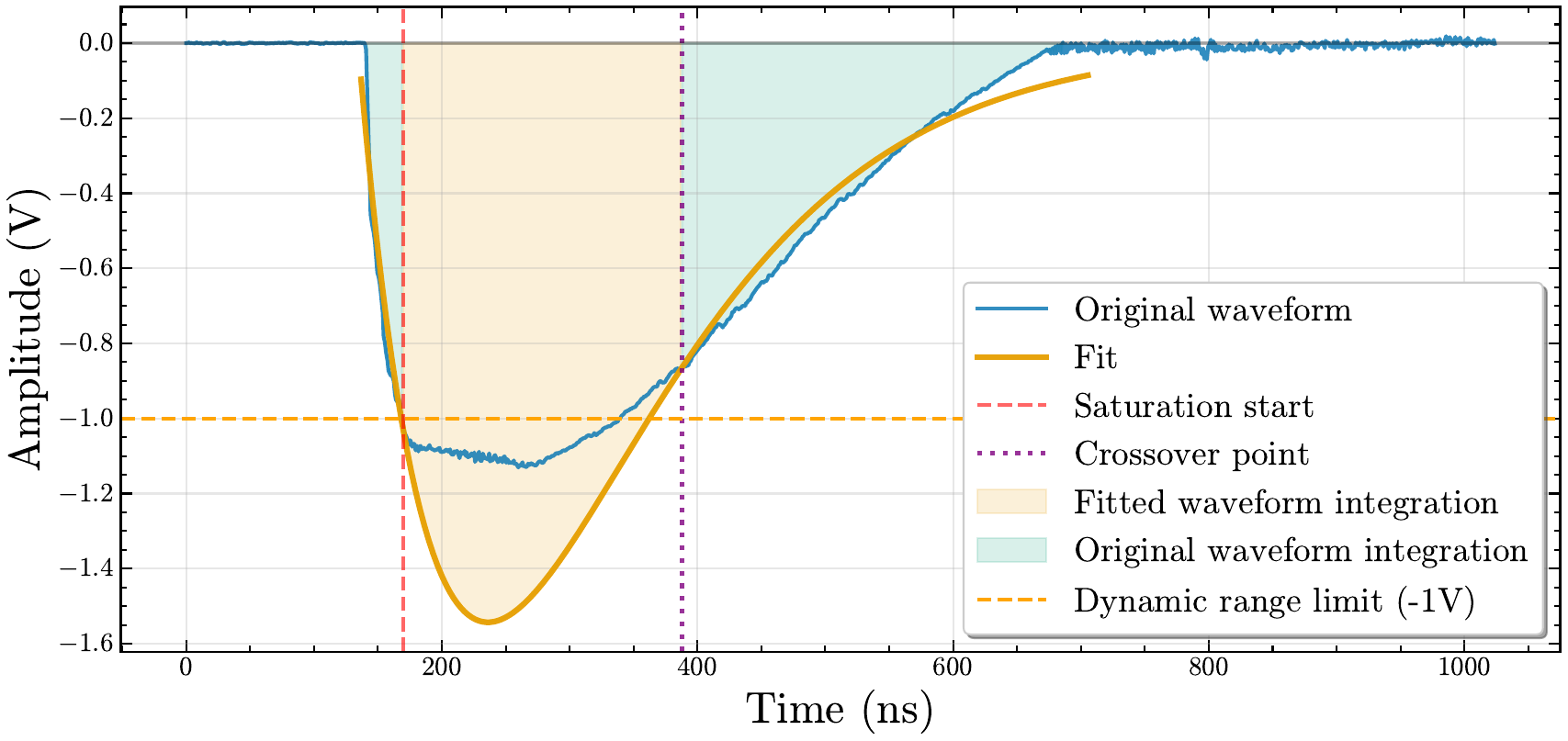}
    \caption{An example pulse fitting for saturation correction in SiPM waveforms from gamma-source data. The measured waveform (blue) exhibits saturation at the digitizer's dynamic-range limit (\SI{-1}{\volt}); the fitted reconstruction (orange) extends beyond the saturation threshold to recover the estimated true pulse shape. The saturated portion of the pulse is reconstructed using the fitted function (orange shading), while the unsaturated portions are integrated from the original waveform (green shading).}
    \label{fig:sat_wf_fit}
\end{figure}

The pulse-shape model used for saturation correction describes the temporal response of photodetector signals for scintillation events and represents a modified version of the Single Avalanche Response Function (SARF) adapted from Ref~\cite{Gallina:2019fxt}. Our modification employs a single effective decay constant to represent the convolution of the intrinsic scintillation decay constant with SiPM-specific recharging processes. The modified SARF model is expressed as:
\begin{equation}
V(t) = -\frac{A}{\tau_{\mathrm{D}}} \left[ \exp\left(-\frac{t-t_0}{\tau_{\mathrm{D}} + \tau_{\mathrm{R}}}\right) - \exp\left(-\frac{t-t_0}{\tau_{\mathrm{R}}}\right) \right],
\end{equation}
\noindent
where $A$ is the pulse amplitude, $t_0$ is the pulse onset time relative to the start of the waveform window, $\tau_{\mathrm{D}}$ is the effective decay time constant including both scintillation decay and SiPM recharging process, and $\tau_{\mathrm{R}}$ is the rise time constant of the SiPM response. Positive function values are clipped to zero to maintain physical consistency.
% , as the pulse template extends to opposite polarity for $t< 0$. 

To correct for signal saturation at the dynamic range limit, pulse fitting was applied to reconstruct the true pulse amplitudes, as demonstrated in Figure~\ref{fig:sat_wf_fit}. Only unsaturated portions of the waveform were used for fitting, and this saturation correction was applied exclusively to gamma events and channels that exceeded the dynamic range. For these corrected pulses, the total charge was calculated using a hybrid integration method combining three regions: the original waveform from the leading edge to the saturation threshold, the fitted waveform through the saturated region until the crossover point where it drops below the recovering original waveform (identified after the pulse peak), and the original waveform from the crossover point to the pulse end. This approach accurately reconstructs the charge in saturated regions while preserving measured data elsewhere, as shown in Figure~\ref{fig:sat_wf_fit}.

%% file: section-sim_fram.tex
% \section{Photon Transport Simulation}\label{sec:chroma_simu_fram}
\section{Optical Simulation Framework}\label{sec:chroma_simu_fram}

To model the detector response and estimate the effective PDE for the LoLX detector, a multi-stage simulation pipeline was developed. The pipeline begins with \textsc{Geant4} \cite{GEANT4:2002zbu} to simulate particle interactions within a geometry that replicates the experimental apparatus, including the stainless steel vacuum chamber, vacuum jacket and LXe volume. Energy depositions from gamma rays within the LXe, designated as the sensitive volume, are recorded and subsequently used as input to the Noble Element Simulation Technique (NEST) \cite{NEST}. NEST generates scintillation photons based on the energy-dependent light yield of LXe, incorporating both Fano factor fluctuations and recombination probability variations to model the stochastic nature of the process \cite{Lenardo:2014cva}.

The subsequent optical photon transport is performed using \textsc{Chroma} \cite{seibert2011chroma}, a GPU-accelerated Monte Carlo ray-tracing framework. This tool was selected for its significant computational efficiency, with photon propagation speeds up to 200 times faster than conventional optical simulations in \textsc{Geant4}. In \textsc{Chroma}, the detector geometry is defined separately from \textsc{Geant4} using triangulated surface meshes from STL files. The framework fully simulates optical processes such as absorption, scattering, and Fresnel reflection. Its customizable photon detection logic was utilized to implement the wavelength- and angular-dependent PDE model detailed in Ref.~\cite{Croix:2025axo}.
% it has been mention below

\begin{figure}
    \centering
    \includegraphics[width=\linewidth]{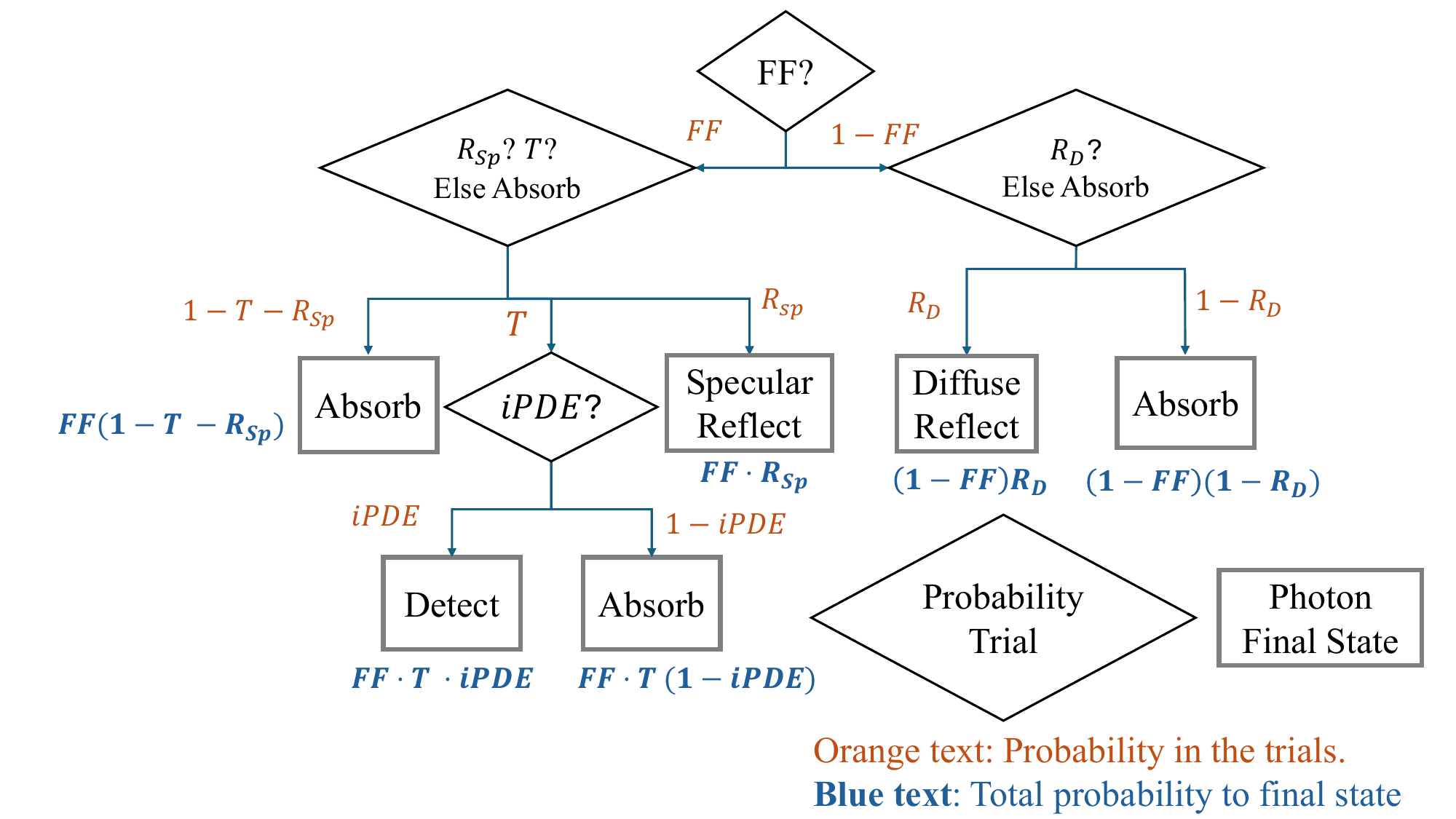}
    \caption{ Flowchart of the Bernoulli trial photon detection scheme for SiPM simulation in \textsc{Chroma}. The decision tree illustrates the sequential probabilistic processes governing photon interactions with SiPMs, including the fill factor $FF$, wavelength and angular dependent transmission through oxide layers into bulk $T(\lambda, \theta)$, specular reflection $R_{Sp}(\lambda, \theta)$, diffuse reflection $R_D$ from inactive regions, and internal photon detection efficiency $iPDE(\lambda, OV)$. Orange text indicates individual trial probabilities at each decision point, while blue text shows the total probability for reaching each final state.}
    \label{fig:chroma_PDE}
\end{figure}

The SiPM efficiency model \cite{Croix:2025axo} in \textsc{Chroma} incorporates vacuum efficiency measurements \cite{Lewis:2024hzs} and xenon optical properties \cite{Grace:2015yta} to accurately model detector response in liquid xenon environments. Figure~\ref{fig:chroma_PDE} illustrates the photon-detection sequence implemented in \textsc{Chroma} as a series of Bernoulli trials. The detection process depends on several parameters, including the fill factor \(FF\), the transmission probability \(T(\lambda,\theta)\), the specular reflectivity \(R_{\mathrm{Sp}}(\lambda,\theta)\), the diffuse reflectivity \(R_D\) from inactive regions, and the internal photon detection efficiency \(iPDE(\lambda,V_{\mathrm{OV}})\). Here, \(T\) and \(R_{\mathrm{Sp}}\) both depend on the wavelength \(\lambda\) and angle of incidence \(\theta\), while \(iPDE\) depends on the wavelength and overvoltage \(V_{\mathrm{OV}}\), and represents the probability that a transmitted photon generates an avalanche. 
The overall $PDE$ of a SiPM is defined as:
\begin{equation}
    PDE(\lambda, \theta, V_{\mathrm{OV}}) = FF\cdot T(\lambda, \theta) \cdot iPDE(\lambda, V_{\mathrm{OV}}). 
    \label{eq:PDE}
\end{equation}

The simulation proceeds through a sequence of probabilistic trials. First, the fill factor test determines whether a photon hits the active SiPM surface or an inactive region. Consistent with the treatment in LoLX 1 \cite{lolx_ext_2025}, the SiPM diffuse reflectivity is assumed to originate solely from the inactive fraction of the sensor. The total diffuse reflectivity measured in \cite{Lv:2019res} is equated to $(1-FF)R_D$. Consequently, we use $R_D=0.45$ for both FBK and HPK, which corresponds to an overall diffuse reflectivity of 0.18 for HPK and 0.10 for FBK, matching the measurements in \cite{Lv:2019res}. For incidence on the photosensitive SiPM surface, the model evaluates for transmission $T$ or specular reflectivity $R_{sp}$, with values taken from lookup tables produced from \cite{Croix:2025axo}. Note that all HPK SiPMs include a \SI{0.5}{mm} thick quartz window, which is modeled in the \textsc{Chroma} geometry and treated separately from the PDE modeling discussed here. In this study, only wavelengths above 160~nm are used, thus $T = 1 - R_{Sp}$ as there is no absorption in the silicon-dioxide film. However, the simulation framework can accommodate $R+T<1$, in which case absorption is defined as $A = 1 - R - T$. Successfully transmitted photons undergo a final detection trial based on the internal efficiency, where they are either converted to a signal with probability $iPDE$ or absorbed without signal generation. Separating the internal PDE from the optical terms allows the overvoltage-dependent detection efficiency to be varied in the offline analysis without modifying the optical transport simulation.

\begin{figure}[htbp]
    \centering
        \includegraphics[width=0.8\linewidth]{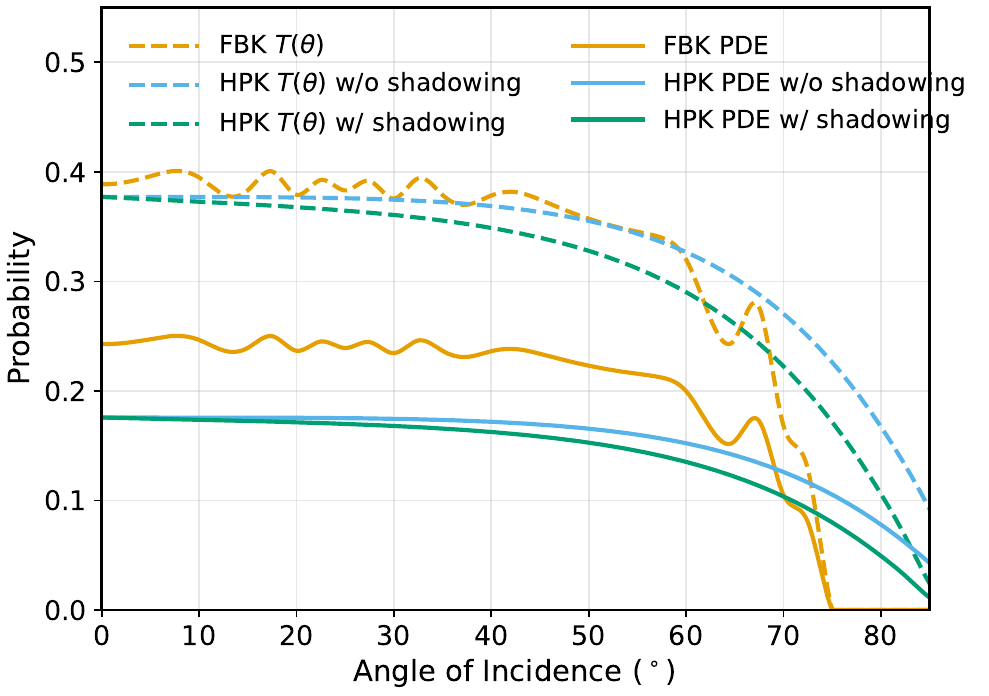}
        \caption{ Angular dependence of optical parameters for FBK~HD3 and HPK~VUV4 SiPMs, modeled at $\lambda=175\,$nm and $V_{\mathrm{OV}}=3\,$V. The plot shows the transmission $T(\theta)$ and the resulting PDE as a function of the incident angle $\theta$ (measured with respect to the normal).} 
        \label{fig:SiPM_PDE_model}
\end{figure}

Figure~\ref{fig:SiPM_PDE_model} shows the angular dependence of transmission, which contributes to the PDE given in Eq.~\eqref{eq:PDE}. Because the FBK~HD3 and HPK~VUV4 share nearly identical transmission and reflection profiles at the xenon scintillation wavelength (\SI{175}{\nano\meter}), the offset between their $PDE(\theta)$ curves is governed almost entirely by their differing geometric fill factors: 0.6 for HPK and 0.8 for FBK.
The transmission for the FBK devices approaches zero at about 75$^\circ$, as shown in Figure \ref{fig:SiPM_PDE_model}. This is a result of thin-film interference effects, which are highly dependent on the thickness of the surface layer. The thicker silicon dioxide layer on the FBK SiPMs alters the conditions for constructive interference, causing high reflectivity to be achieved at shallower grazing angles compared to the HPK SiPMs.
The simulation also includes an additional correction factor suggested in \cite{Croix:2025axo}, where the HPK SiPM exhibits shadowing of the active region at high incidence angles due to surface structures. This correction factor is applied to the angular-dependent transmission and reflectivity, which impacts photon transport and the detector's effective PDE. 

The breakdown voltage was measured with \SI{0.5}{V} sampling steps in the experiment, resulting in a systematic uncertainty of $\pm$\SI{0.5}{V} in the nominal operating voltage.
The simulation incorporated the same systematic uncertainties affecting the experimental measurements. The overvoltage uncertainty of $\pm\SI{0.5}{\volt}$ was propagated through the simulation by varying the internal PDE model parameters within this range. The crosstalk effects were simulated independently based on the external crosstalk model in \cite{lolx_ext_2025} to estimate their impact on the photon detection ratio.
% External crosstalk photons have wavelengths distributed from \num{600} to \SI{1000}{\nano\meter}, and the PDE model in \cite{Croix:2025axo} indicates that HPK exhibits higher efficiency at longer wavelengths. Consequently, HPK SiPMs are more sensitive to external crosstalk photons than FBK SiPMs.
Two major uncertainties are included in the final results.

%% file: section_SiPMCharge.tex
\subsection{Experimental Comparison of SiPM Response to LXe Scintillation}\label{sec:experiment}

To compare the performance of FBK and HPK SiPMs in liquid xenon, external $^{137}$Cs (\SI{662}{keV}) and $^{133}$Ba gamma sources have been used to generate scintillation light in LoLX. The $^{133}$Ba source emits gamma rays at several energies, with its most prominent line at \SI{356}{keV} \cite{NNDC}. Before data-taking, the gain of the detector is balanced by measuring the breakdown voltage for each channel to supply a constant overvoltage for each device. Channel-specific SPE calibration allows for estimation of the number of PE in each SiPM and PMT channel. This enables a direct comparison of the measured scintillation light yield between photosensor groups.

\begin{figure}[htbp]
    \centering
    % --- TOP ROW: EXPERIMENT ---
    \begin{subfigure}[b]{0.49\linewidth}
        \centering
        \includegraphics[width=\linewidth]{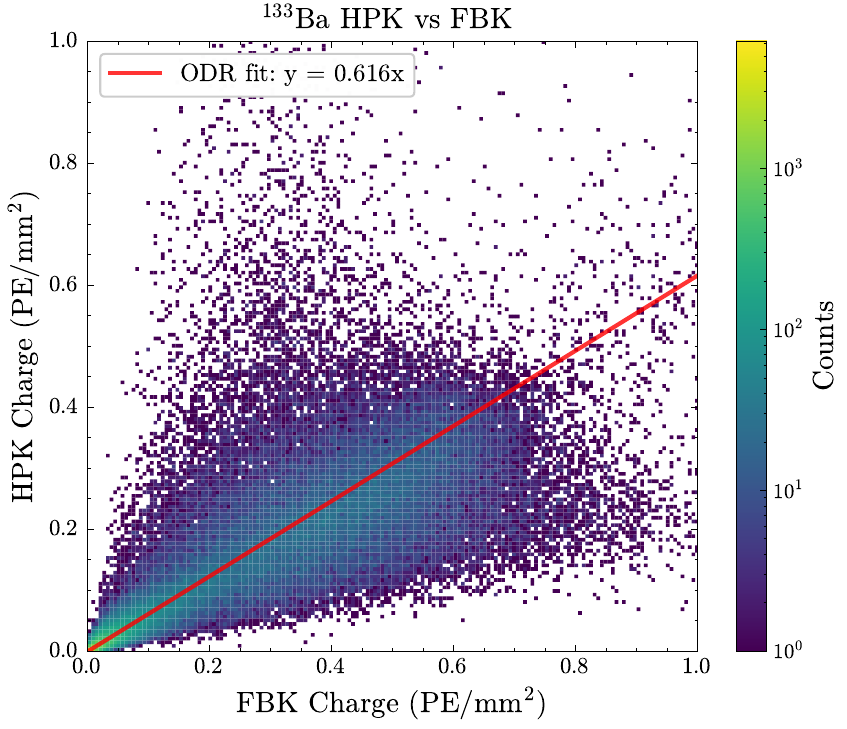} % Replace with your file
        \caption{Experimental ${}^{133}\text{Ba}$}
        \label{fig:Ba_Exp}
    \end{subfigure}
    \hfill
    \begin{subfigure}[b]{0.49\linewidth}
        \centering
        \includegraphics[width=\linewidth]{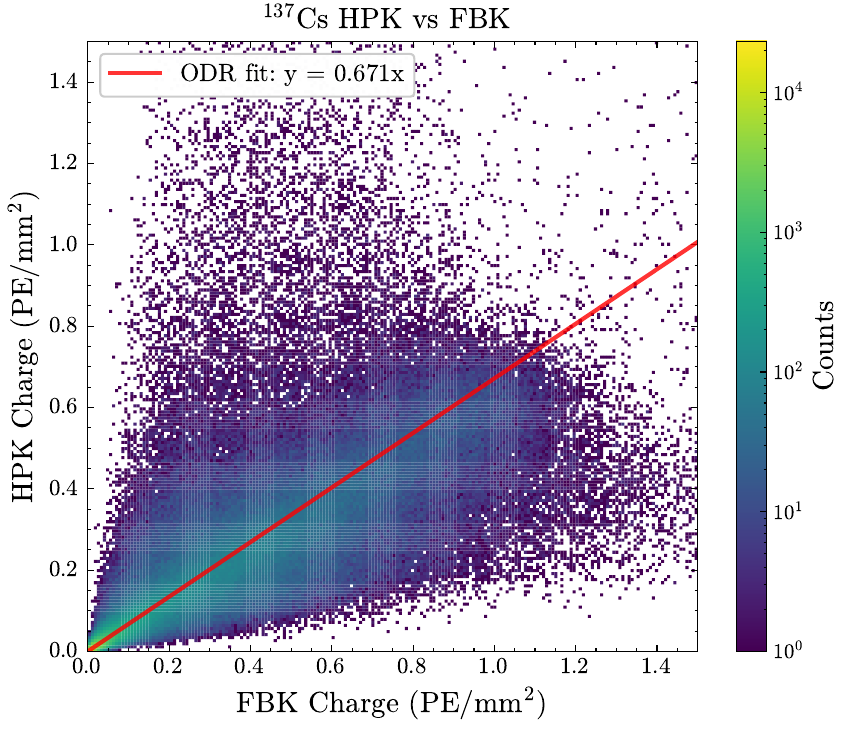} % Replace with your file
        \caption{Experimental ${}^{137}\text{Cs}$}
        \label{fig:Cs_Exp}
    \end{subfigure}
    
    % --- BOTTOM ROW: SIMULATION ---
    \vspace{0.2cm} % Add a little vertical space between rows
    
    \begin{subfigure}[b]{0.49\linewidth}
        \centering
        \includegraphics[width=\linewidth]{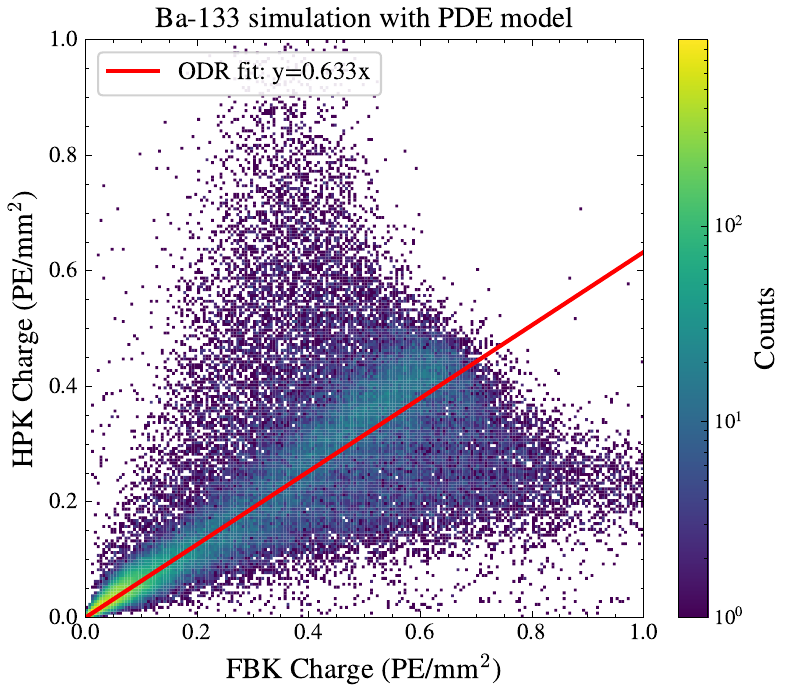} % Your old Fig 6b
        \caption{Simulated ${}^{133}\text{Ba}$}
        \label{fig:Ba_Sim}
    \end{subfigure}
    \hfill
    \begin{subfigure}[b]{0.49\linewidth}
        \centering
        \includegraphics[width=\linewidth]{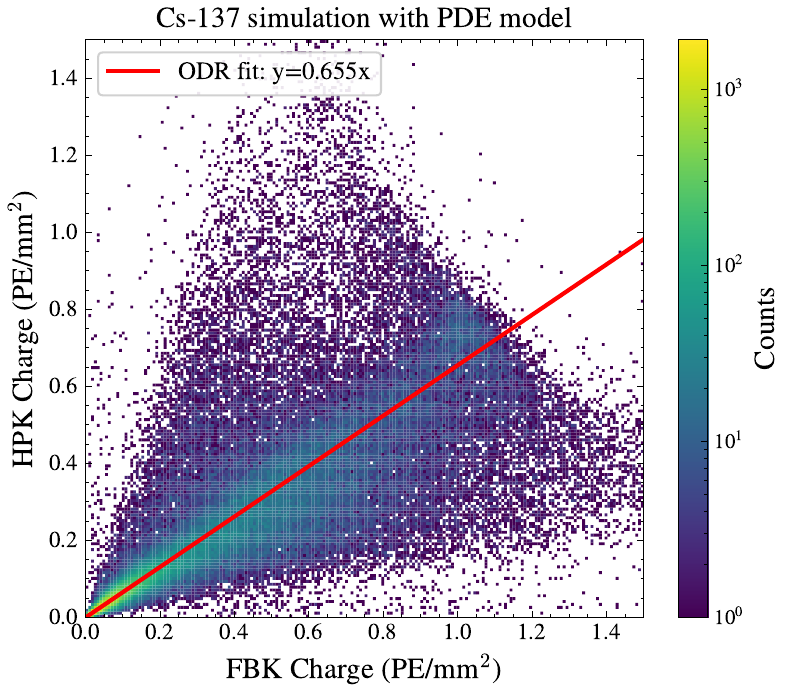} % Your new Cs plot
        \caption{Simulated ${}^{137}\text{Cs}$}
        \label{fig:Cs_Sim}
    \end{subfigure}
    
    \caption{Comparison of measured and simulated charge per unit area for HPK~VUV4 versus FBK~HD3 SiPMs. 
    Top Row: Experimental data for (a) ${}^{133}\text{Ba}$ and (b) ${}^{137}\text{Cs}$ interactions. 
    Bottom Row: Simulated data for (c) ${}^{133}\text{Ba}$ and (d) ${}^{137}\text{Cs}$ using the angular-dependent PDE model with geometric shadowing effects included. The red lines show orthogonal distance regression (ODR) fits.
     }
    \label{fig:HPK_vs_FBK_Comparison}
\end{figure}

Figure~\ref{fig:HPK_vs_FBK_Comparison} shows the direct comparison between the two types of SiPMs by plotting a 2D distribution, representing the pulse charge estimation detected by FBK devices versus HPK devices.
The total charge measured by each SiPM vendor group is normalized by the
corresponding operational sensor area. The area is calculated as the number of operational sensors multiplied by the nominal \(6~\mathrm{mm}\times 6~\mathrm{mm}\) surface area of each sensor. This normalization accounts for disabled or non-operational channels but does not correct for the SPAD fill factor within each SiPM.

The red line represents an orthogonal distance regression (ODR) linear fit to the data, revealing the correlation between charge detection in the two sensor types. ODR was used because both detector types (shown on either axis) have similar measurement uncertainties, arising from position variation and charge reconstruction of the external source.

The fitted slopes of $0.62^{+0.03}_{-0.04}$ for $^{133}\text{Ba}$ and $0.67^{+0.03}_{-0.05}$ for $^{137}\text{Cs}$ indicate that the HPK SiPMs detect fewer photons than the FBK devices.  We observe that FBK sensors achieve a higher overall photon detection under the same experimental conditions when compared to the HPK VUV-4 sensors.
While this difference in detection is partly explained by the lower fill factor of the HPK SiPMs, the measured in-situ ratio remains notably lower than corresponding vacuum normal-incidence measurements, as will be detailed in the Discussion Section~\ref{sec:discussion}.
The uncertainties in the fitted slopes incorporate systematic errors from SPE calibration and overvoltage uncertainties on the devices.

%% file: section-sim_results.tex
\subsection{Comparison with Optical Simulations
}\label{sec:sim_result}

To validate our model against experimental data, we performed simulations with  \({}^{133}\mathrm{Ba}\) and \({}^{137}\mathrm{Cs}\) external sources, following  the complete pipeline described in Section~\ref{sec:chroma_simu_fram}. We compared the photon detection rates between HPK and FBK SiPMs, calculating their ratio analogous to the experimental analysis shown in Figure~\ref{fig:HPK_vs_FBK_Comparison}. All simulations were performed at an overvoltage of $V_{\mathrm{OV}} = \SI{3}{\volt}$ to match the experimental conditions. Figure~\ref{fig:Ba_Sim} and Figure~\ref{fig:Cs_Sim} present the simulation results comparing light detection between HPK and FBK SiPMs.

Sources of systematic uncertainty include the $\pm$\SI{0.5}{V} uncertainty in overvoltage and the external crosstalk. External crosstalk photons have wavelengths distributed from \num{600} to \SI{1000}{\nano\meter}, and the PDE model in \cite{Croix:2025axo} indicates that HPK exhibits higher efficiency at longer wavelengths. Consequently, HPK SiPMs are more sensitive to external crosstalk photons than FBK SiPMs, thereby affecting the upper bound of the photon detection ratio. 
However, this effect increases the photon detection ratio by only about \SI{3}{\%} relative to the value obtained without external crosstalk. After propagating these two major systematic uncertainties, the simulation results with their associated uncertainties are presented in Table~\ref{tab:detection_ratio}.

\begin{table}[htbp]
  \centering
      \caption{ 
            Comparison of HPK-FBK photon detection ratios at $V_{\mathrm{OV}}=\SI{3}{\volt}$ and $\lambda \approx \SI{175}{\nano\meter}$.
            These experimental and simulated values represent the effective in-situ detection ratio, which accounts for both the intrinsic SiPM PDE and the geometric transport efficiency.
            The simulations utilize an angular-dependent SiPM model from Ref.~\cite{Croix:2025axo}, and are compared against both a normal-incidence model value and a vacuum measurement~\cite{nEXO:2022yey}.
            } \label{tab:detection_ratio}
  \begin{tabular}{l c c}
    \toprule\toprule % Top rule
    Method & \multicolumn{2}{c}{HPK/FBK Photon Detection Ratio} \\
    \cmidrule(lr){2-3} % Partial rule under columns 2-3
    & $^{133}$Ba & $^{137}$Cs \\
    \midrule % Middle rule
    Vacuum measurement ($0^{\circ}$ AOI)~\cite{nEXO:2022yey} & \multicolumn{2}{c}{$0.84 \pm 0.01$} \\
    PDE model extrapolation in LXe ($0^{\circ}$ AOI) ~\cite{Croix:2025axo} & \multicolumn{2}{c}{$0.72 \pm 0.03$} \\
    \midrule % Rule to separate experimental from simulation
    \textbf{Experimental Data (This work)} & & \\
    \textit{LoLX2 data} & $0.62^{+0.03}_{-0.04}$ & $0.67^{+0.03}_{-0.05}$ \\
    \midrule % Rule to separate sections
    \textbf{Simulation Model} & & \\
    \textit{Full simulation (with HPK shadowing)} & $0.63^{+0.03}_{-0.04}$ & $0.66^{+0.04}_{-0.05}$ \\
    \textit{Simulation w/o HPK shadowing} & $0.68^{+0.04}_{-0.05}$ & $0.71^{+0.04}_{-0.05}$ \\
    \textit{Full + excess absorption} & $0.55^{+0.03}_{-0.04}$ & $0.58^{+0.03}_{-0.04}$ \\
    \bottomrule\bottomrule % Bottom rule
  \end{tabular}

\end{table}

As summarized in Table~\ref{tab:detection_ratio}, the experimental HPK-FBK photon detection ratios for LXe scintillation from ${}^{133}\mathrm{Ba}$ ($0.62^{+0.03}_{-0.04}$) and ${}^{137}\mathrm{Cs}$ ($0.67^{+0.03}_{-0.05}$) are in good agreement with our full simulation predictions, which include surface structural shadowing effects for the HPK SiPMs ($0.63^{+0.03}_{-0.04}$ for ${}^{133}\mathrm{Ba}$ and $0.66^{+0.04}_{-0.05}$ for ${}^{137}\mathrm{Cs}$). Quantitatively, the differences between the measured and simulated ratios correspond to $0.20~\sigma$ ($p=0.84$) for ${}^{133}\mathrm{Ba}$ and $0.16~\sigma$ ($p=0.88$) for ${}^{137}\mathrm{Cs}$. By contrast, the simulation without HPK shadowing predicts systematically higher ratios ($0.68^{+0.04}_{-0.05}$ and $0.71^{+0.04}_{-0.05}$), corresponding to larger deviations of $1.03~\sigma$ ($p=0.30$) for ${}^{133}\mathrm{Ba}$ and $0.69~\sigma$ ($p=0.49$) for ${}^{137}\mathrm{Cs}$. This demonstrates that surface shadowing is a non-negotiable effect that must be included to accurately model the detector response.

%% file: section-Discussion.tex
\subsection{Discussion}\label{sec:discussion}

\begin{figure}[htbp]
    \centering
    \includegraphics[width=0.8\linewidth]{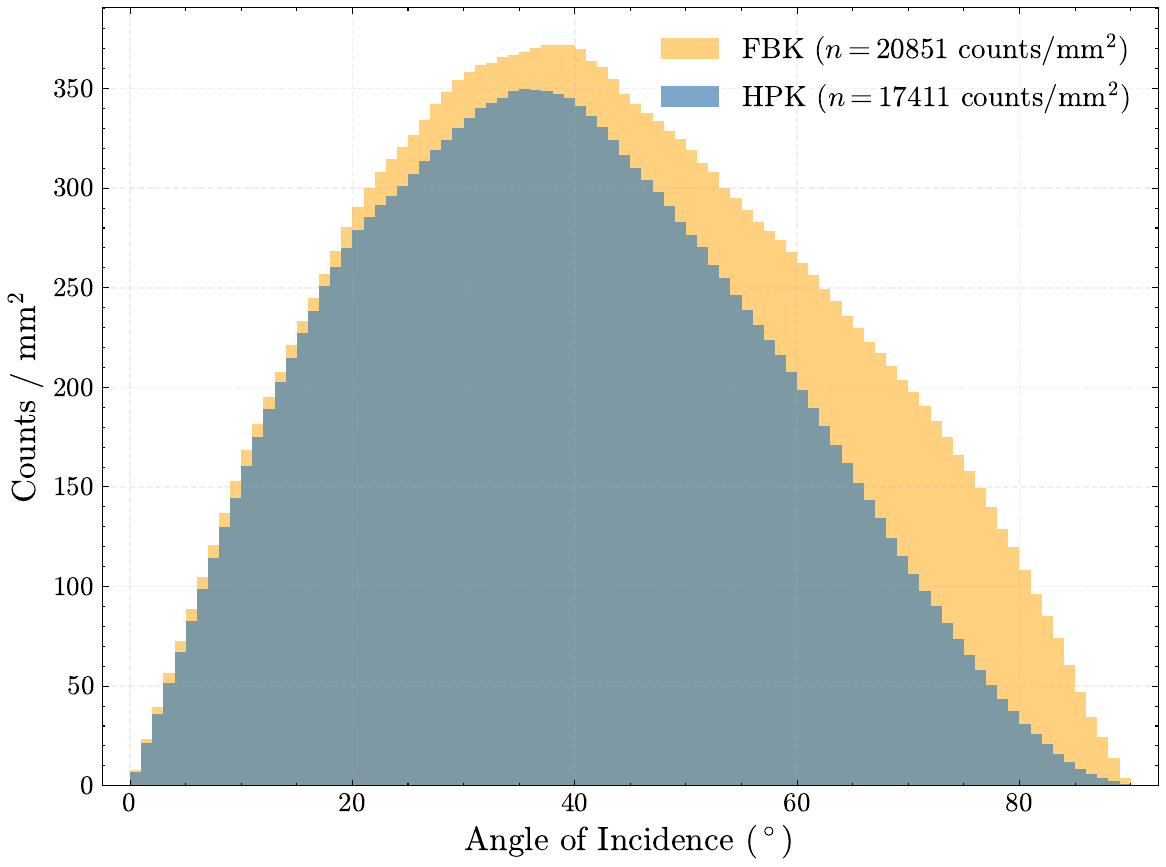}
    
    \caption{AOI distribution for photons at their first hit with the SiPM in a $^{137}$Cs simulation, normalized by the active SiPM area. For the photons arriving at large angles; they are subject to both the surface shadowing shown in Figure~\ref{fig:SiPM_PDE_model} and additional macroscopic shadowing from the HPK package geometry (ceramic walls), which reduces the photon transport efficiency.}
    \label{fig:SiPM_AOI_dist}
\end{figure}

Detailed analysis of the simulation allows us to decouple the intrinsic sensor PDE from macroscopic geometric effects. The photon transport efficiency is defined as the probability that a generated photon successfully hits the active sensor area. The simulation reveals that this transport efficiency is approximately 16\% lower for HPK VUV4 devices compared to FBK HD3 (a ratio of $\sim0.84$). This deficit is primarily caused by the macroscopic HPK package geometry (ceramic walls and window), which effectively block the large population of grazing-incidence photons shown in Figure~\ref{fig:SiPM_AOI_dist}. When this geometric transport factor ($0.84$) is combined with the intrinsic angular photon detection ratio, it fully accounts for the observed effective photon detection ratio of $0.62$--$0.67$.
 Geometric effects also explain why the measured photon detection ratio is higher for the \({}^{137}\text{Cs}\) source. The higher-energy \(\SI{662}{keV}\) gammas from \({}^{137}\text{Cs}\) have a longer mean free path in LXe than the \(\SI{356}{keV}\) gammas from \({}^{133}\text{Ba}\), leading to a more spatially uniform distribution of energy throughout the detector volume. This uniformity results in an average photon Angle of Incidence (AOI) distribution that is slightly closer to normal incidence, yielding a higher effective photon detection ratio that approaches the normal-incidence ratio of \(0.72 \pm 0.03\) from Ref.~\cite{Croix:2025axo}.

Independent measurements from Ref.~\cite{Lv:2019res} indicate that the specular reflectivity of HPK VUV4 devices is 20--25\% lower than the data used in the PDE model~\cite{Croix:2025axo}. This discrepancy is also visible in Figure~9 of Ref.~\cite{Croix:2025axo}, where the vacuum reflectivity data from Ref.~\cite{Lv:2019res} lie below the model expectation, while the same HPK optical description, including shadowing, better describes the other angular VUV4 measurements. The origin of this discrepancy is not clear, but possible explanations include device-to-device or batch-to-batch variation, differences in surface condition or contamination, or unaccounted-for systematic effects in the optical measurements. To evaluate the possible impact of this discrepancy, we hypothesize and test the effect of excess absorption in the SiPM surface layers, since specularly reflected photons traverse these layers twice. We artificially included this enhanced absorption in our simulation by reducing the transmission by \SI{12.5}{\%} and the specular reflectivity by \SI{25}{\%}, which decreases the predicted HPK-to-FBK detection ratio to \(0.55^{+0.03}_{-0.04}\) for \({}^{133}\mathrm{Ba}\) and \(0.58^{+0.03}_{-0.04}\) for \({}^{137}\mathrm{Cs}\). These values undershoot the experimental data by approximately \(1.3~\sigma\) and \(1.7~\sigma\), respectively. Therefore, while absorption in the surface layers may contribute to differences between optical datasets or devices, our data do not favor applying a reflectivity reduction of the magnitude reported in Ref.~\cite{Lv:2019res}.

These findings highlight that the effective photon detection ratio is a convolution of the intrinsic angle dependent efficiency of the SiPMs and system-level photon transport determined by detector geometry and event distributions. Thus, accurate detector simulation requires a full optical simulation integrating an angular-dependent PDE model. Ultimately, the relevant efficiency metric is not the peak normal-incidence efficiency, but this holistic treatment that includes the distribution of photon arrival angles.

%% file: section-conclusion.tex
\section{Conclusion}\label{sec:conclusion}

In this work, the relative photon detection of FBK VUV-HD3 and HPK VUV4 SiPMs in the LoLX 2 liquid xenon detector was measured using external \({}^{133}\text{Ba}\) and \({}^{137}\text{Cs}\) gamma sources. The experimental HPK-FBK photon detection ratios were found to be \(0.62^{+0.03}_{-0.04}\) for \({}^{133}\text{Ba}\) and \(0.67^{+0.03}_{-0.05}\) for \({}^{137}\text{Cs}\). These results indicate that, within this specific detector geometry, the photon detection ratio of the HPK SiPMs is \num{33}--\SI{38}{\%} lower than that of the FBK sensors. This difference is driven by the combination of the HPK devices' lower fill factor and the significant impact of structural shadowing and package shadowing, which is most pronounced for photons arriving at low grazing angles.

Our full optical simulation, which incorporates an angular-dependent PDE model and a correction for surface shadowing effects, successfully reproduces these experimental ratios. This agreement validates our optical transport framework and confirms that detailed geometric and surface effects are critical for accurate performance predictions. The measured effective photon detection ratios (0.62–0.67) are substantially lower than the normal-incidence vacuum measurement (0.84). This result validates the effectiveness of the PDE model. Consequently, future large-scale rare-event searches cannot rely on normal-incidence specifications alone; robust characterization requires system-level simulations that convolve the angular-dependent PDE with the full optical transport of scintillation light.